\documentclass[twocolumn,twoside,slac_two]{revtex4}
\usepackage{graphicx}
\usepackage{bm}
\usepackage{fancyhdr}
\def\wf{\widetilde{F}^{(0,3)}}
\begin{document}

\title{How dark matter cares about topological superstrings}

%

\author{Luis A. Anchordoqui}
\affiliation{Department of Physics, University of Wisconsin-Milwaukee,  Milwaukee, WI 53201, USA}

\begin{abstract}
Non-trivial toplogical properties of string world-sheets with three boundaries can give rise to superpotentials which preserve supersymmetry but violate R-symmetry by two units. This results in four point functions which permit s-wave annihilation of two neutralinos into gauge bosons. If the topological partition function is such as to allow saturation of the WMAP dark matter density for low string scales ($M_s \sim 2$~TeV), the annihilation into monochromatic gamma rays is predicted to lie about a factor of 2 below the current H.E.S.S. measurement of gamma ray flux from the galactic center. Thus, it may be detectable in the next round of gamma ray observations.
\end{abstract}

\maketitle

\thispagestyle{fancy}


\section{Setting the stage}

Cosmological and astrophysical observations provide plentiful evidence
that a large fraction of the universe's mass consists of non-luminous,
non-baryonic material, known as dark
matter~\cite{Bertone:2004pz}.  Among the
plethora of dark matter candidates, weakly interacting massive
particles (WIMPs) are especially well-motivated, because they combine
the virtues of weak scale masses and couplings, and their stability
often follows as a result of discrete symmetries that are mandatory to
make electroweak theory viable (independent of cosmology). Indeed,
there are tantalizing clues of the connection between electroweak
symmetry breaking and dark matter.  Given electroweak-strength
interactions between WIMPs and the Standard Model, the thermal
freeze-out abundance of dark matter can be roughly related to the
thermally-averaged WIMP annihilation cross section, $\langle \sigma v
\rangle$, through $\Omega h \sim 10^{-10}~{\rm GeV}^{-2}/\langle
\sigma v \rangle$~\cite{Jungman:1995df}.  A typical
weak cross section is $\langle \sigma v \rangle \sim \alpha^2/M_{\rm
  weak}^2 \sim 10^{-9}~{\rm GeV}^{-2}$. This corresponds to a thermal relic density $\Omega h^2 \sim 0.1$, which agrees with cosmological observations $\Omega h^2 = 0.113 \pm 0.003$~\cite{Komatsu:2008hk}. Of course, to expose the identity of dark matter, it is necessary to measure its non-gravitational couplings. Efforts in this direction include direct detection experiments, which hope to observe the scattering of dark matter particles with the target material of the detector, and indirect detection experiments which are designed to search for the products of WIMP annihilation into gamma-rays, anti-matter, and neutrinos.

The galactic center (GC) has long been considered to be among the most
promising targets for detection of dark matter annihilation,
particularly if the halo profile of the Milky Way is cusped in its
inner volume~\cite{Navarro:1996gj}.  However, a major adjustment in
the prospects for indirect dark matter detection has materialized
recently, following the discovery of a bright astrophysical source of
TeV gamma-rays at the
GC~\cite{Aharonian:2004wa}. This implies that dark
matter emission from the GC will not be detectable in a (quasi)
background-free regime, and---unless one focuses attention to other
targets---the peculiar spectral shape and angular distribution of dark
matter annihilation must be used to isolate the signal from
background.

An attractive feature of broken SUSY is that with R-parity conservation, the lightest supersymmetric particle ($\chi^0$) becomes an appealing dark matter candidate~\cite{Goldberg:1983nd}.\footnote{R-parity is defined by $R_p = (-1)^{3(B-L)+2S},$ where $B$, $L$, and $S$ are the baryon number, lepton number, and spin, respectively. All Standard Model  particles have $R_p=1$ and all superpartners have $R_p =-1$.  Conservation of R-parity implies $\Pi R_p =1$ at each vertex, yielding an elegant way to forbid proton decay in SUSY models (e.g. in R-parity conserving models $p \to \pi^0 e^+$ cannot be mediated by a squark).}  In particular, neutralinos annihilating in the halo of the Milky Way to final states containing a photon (such as $\gamma \gamma$ or $\gamma Z$) lead to a very distinctive gamma ray line which could potentially provide a ``smoking gun'' signature for annihilating dark matter, if sufficiently bright. Unfortunately, the annihilation is hindered because of p-wave barrier. To create an s-wave, both gauginos must be in the same helicity state, either left- or right-handed. This certainly complicates the panorama, because in SUSY the gaugino pair $\lambda^\pm \lambda^\pm$
is in the $A$-term of the chiral superfields $WW$ and the two gauge bosons are in the $F$-term of the same or different $WW|_{\theta \theta}$ pair, which implies that each $WW$ is by itself a superpotential and that both $WW$ have identical R-charges. Then, such gaugino pair annihilation violates R-symmetry by two units ($\Delta r = \pm 2$), and is forbidden in (unbroken) supersymmetric Yang-Mills theory, at least at the perturbative level.\footnote{R-symmetry is a continuous $U(1)$ symmetry under which the 3 members of the chiral superfields transform in a prescribed way: $A \to e^{2in\alpha} A,$ $\psi \to e^{2i (n-\frac{1}{2}) \alpha} \psi$, $F\to e^{2i(n-1) \alpha} F$, where $n$ is called the R-character~\cite{Wess-Bagger}.  Mass terms or potentials are R-invariant only if the R-characters of their respective superfields add up to one. For products of superfields the R-charged deficit $\Delta r$ is the sum of the characters.  There is a discrete subset from R-symmetry that can be connected to R-parity. {\it A priori} one can break the continuous symmetry and leave the subset of the discrete symmetry invariant. In the model we will discuss here the R-symmetry breaking is in the gauge sector.} However, it can appear in conjunction with a SUSY-breaking gaugino mass generation mechanism. In the Minimal Supersymmetric Standard Model (MSSM) the one-loop $\chi^0 \chi^0 \to \gamma \gamma \, (\gamma Z)$ proccesses are dramatically suppressed to a branching fraction of 0.1\%~\cite{Bergstrom:1997fh}.

\section{WIMPs from intersecting D-branes}

In this work we focus attention on intersecting D-brane configurations that realize the Standard Model by open strings~\cite{Blumenhagen:2006ci}.  To develop our program in the simplest way, we work within the construct of a minimal model in which  the (color) $U(3)_a$ stack of D-branes is intersected by the (weak doublet) $U(2)_b$ stack of D-branes, as well by a third (weak singlet) $U(1)_c$ stack of D-brane~\cite{Antoniadis:2000ena}. 
In the bosonic sector, the open strings terminating on the $U(3)_a$ stack contain the standard octet of gluons $g_\mu^a$ and an additional $U(1)_a$ gauge boson $C_\mu$ (most simply the manifestation of a gauged baryon number symmetry); on the $U(2)_b$ stacks the open strings correspond to the electroweak gauge bosons $W_\mu^a$, and again an additional $U(1)_b$ gauge field $X_\mu$.  So the associated gauge groups for these stacks are $SU(3)_C \times U(1)_a,$ $SU(2)_L \times U(1)_b$, and $U(1)_c$, respectively. The $U(1)_Y$ boson $Y_\mu$, which gauges the usual electroweak hypercharge symmetry, is a linear combination of $C_\mu$, the $U(1)_c$ boson $B_\mu$ and $X_\mu$. The fermionic matter consists of open strings, which stretch between different stacks of D-branes and are hence located at the intersection points. The vector boson $Y'_\mu$, orthogonal to the hypercharge, must grow a mass  in order to avoid long range forces between baryons other than gravity and Coulomb forces. The anomalous mass growth allows the survival of global baryon number conservation, preventing fast proton decay.

We consider the introduction of new operators, based on superstring theory, which avoids p-wave suppression by permitting neutralino s-wave annihilation into monochromatic gamma rays at an adequate rate. There is a topological theorem that relates the Euler characteristic of the string world-sheet to the change in R-charge via $|\Delta r | \leq -2 \chi$.  For a gaugino pair to annihilate into gauge bosons one needs a world-sheet with Euler characteristic $\chi= 2 -2g -h = -1$. It can be realized in two ways: a ``genus 3/2'' world-sheet ($g=1,\, h=1$)~\cite{Antoniadis:2004qn}, and a two-loop open string world-sheet ($g=0,\, h=3$)~\cite{Antoniadis:2005sd}. We may  choose a supersymmetric R-symmetry violating effective Lagrangian incorporating the above properties, once gauginos acquire mass through an unspecified mechanism. The topological Lagrangian comes out after summing in a semiclassical way on intermediate winding states~\cite{Anchordoqui:2009bn}
\begin{eqnarray}
\label{lint}
{\cal L}_{\rm eff} & = & \frac{{\cal T}}{M_s^3} \,({\rm Tr}\,
WW)({\rm Tr}\,
WW)\big|_{\theta \theta} + {\rm c.c.} \nonumber \\ 
 & = & \frac{{\cal T}}{8 M_s^3}  \,({\rm Tr}\,
\lambda \lambda)({\rm Tr}\, (FF) + {\rm c.c.} ,
\end{eqnarray}
where $M_s$ is the string mass scale, $W$ are the usual chiral superfields with field strengths associated to appropriate gauge groups, and the traces are taken in the fundamental representations. Here, ${\cal T}=3N\,g_s^3\ \wf $, $g_s$ is the string coupling, and $N$ is the number of D-branes attached to the empty boundary.  (A total of six possibilities in the three-stack model under consideration.) The factor of 3 is the number of choices of the empty boundary.  The factor $F^{(0,3)} = 3N\wf$ is the genus zero topological partition function on a world-sheet with $h=3$ boundaries. It depends on the moduli of compact space and takes into account various string configurations in six internal dimensions.

We can then constrain the free parameters of the model to acquire a neutralino relic density consistent with the measured abundance of dark matter~\cite{Komatsu:2008hk}. With a choice of binos (hypercharge gauge bosons) as our $\chi^0$, and with the assumption of relatively small mixing with the other $U(1)$ subgroups in stacks $a$ and $b$~\cite{Anchordoqui:2007da}, the bino is largely associated with the $U(1)$ stack $c$. At threshold $(s\approx 4m_{\chi^0}^2)$, the total
annihilation rate into gauge bosons must satisfy,
\begin{equation}
\left. \sigma v \right|_{WW} + \left. \sigma v \right|_{gg} +\left. \sigma v \right|_{BB}
\simeq 10^{-9}~{\rm GeV}^{-2} \, ,
\end{equation}
where
\begin{equation}
\left. \sigma v \right|_{WW} =  \frac{3c}{4\pi}\, {\cal T}^2
\left(\frac{\hbar}{M_s\ c}\right)^2  \rho^4  \,,
\label{sigmavWW}
\end{equation}
\begin{equation}
\left. \sigma v \right|_{gg} =  \frac{8c}{4\pi}\, {\cal T}^2
\left(\frac{\hbar}{M_s\ c}\right)^2  \rho^4  \, ,
\label{sigmavgg}
\end{equation}
\begin{equation}
\left. \sigma v \right|_{BB} = \zeta^2 \frac{c}{4\pi}\, {\cal T}^2
\left(\frac{\hbar}{M_s\ c}\right)^2  \rho^4  \, ,
\label{sigmavBB}
\end{equation}
with $\rho\equiv m_{\chi^0}/M_s$.  The factor $\zeta$ parameterizes the uncertainty in the $\chi\chi\rightarrow BB$ amplitude because of non-topological components in the matrix element~\cite{Anchordoqui:2009bn}.  Dominance by the topological component corresponds to $\zeta\approx +1.$ 

There is a wide range of the parameter space satisfying this constraint. In what follows we adopt as fiducial values $g_s =0.2,$ $M_s = 2 $~TeV, $m_{\chi^0} = 1$~TeV, and $\wf = 2.8$. A property inherent to the model is that fixing the total annihilation rate yields  a 10\% branching fraction for $\chi^0\chi^0 \to \gamma \gamma$~\cite{Anchordoqui:2009bn}. 
For neutralinos with masses above a few hundred GeV, H.E.S.S.'s observations of the  GC~\cite{Aharonian:2004wa} can be used to probe the dark matter's annihilation cross section. It is this that we now turn to study.

\begin{figure*}[t]
\centering
\includegraphics[width=135mm]{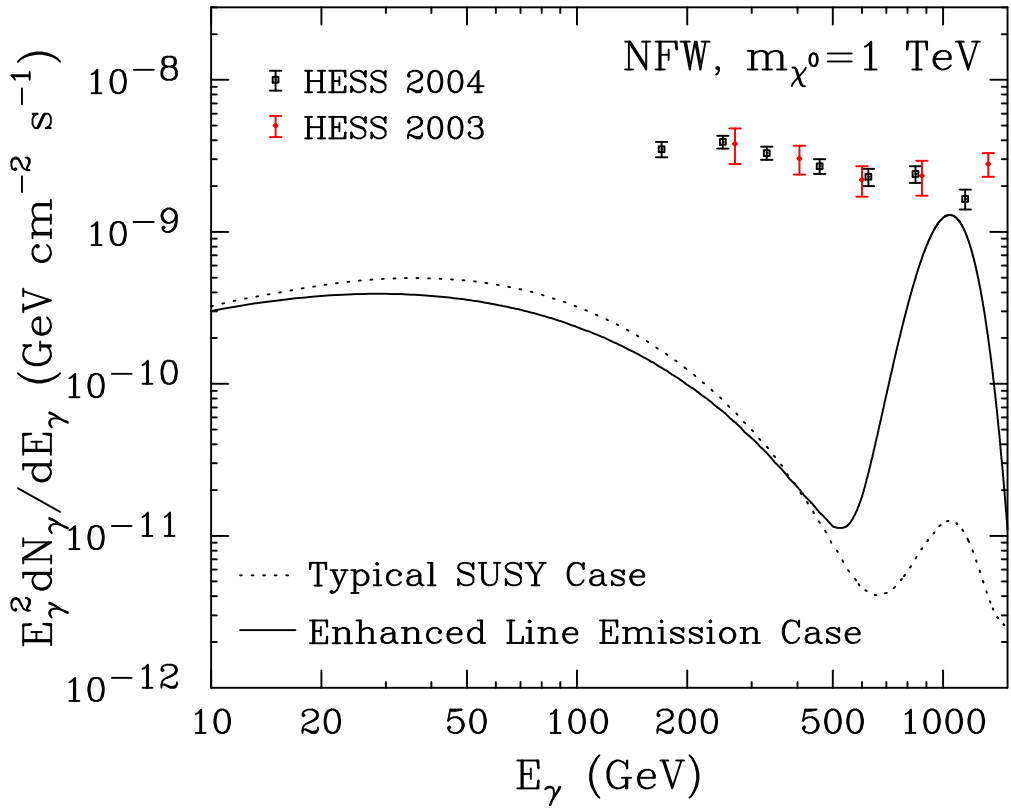}
\caption{ Gamma ray spectrum from neutralino dark matter
  annihilating in the GC  (within a solid angle of
  $10^{-3}$~sr). The spectrum has been convolved with a gaussian of $\Delta
  E_\gamma/E_\gamma$ =15\% width, the typical energy resolution of
  H.E.S.S. and other ground based gamma ray telescopes. The continuum portion of the spectrum arises from the decay
  products of the W and Z bosons and QCD gluons as calculated using Pythia. Also
  shown for comparison are the measurements from
  H.E.S.S.~\cite{Aharonian:2004wa} which are generally interpreted to
  be of astrophysical origin~\cite{Aharonian:2004jr}.}
\label{fig3}
\end{figure*}

\section{Monochromatic $\bm{\gamma}$-rays from the GC}

The differential flux of photons arising from dark matter annihilation observed in a given direction making an angle $\psi$
with the direction of the GC is given
by~\cite{Bergstrom:1997fj}
\begin{equation}
\phi^{\gamma}  = \int \bar{J} \  \frac{1}{2} \ \frac{D_{\odot}}{4\pi} \ \frac{\rho^2_{\odot}}{m_\chi^2} \ \sum_f \, \langle \sigma v \rangle_f \ \frac{dN_f}{dE_{\gamma}} \, d\Omega, 
\label{flux} 
\end{equation}
where $\bar{J} = (1/\Delta \Omega) \int_{\Delta \Omega}
J(\psi)\,d\Omega$ denotes the average of $J$ over the solid angle
$\Delta \Omega$ (corresponding to the angular resolution of the
instrument) normalized to the local density: $J(\psi) =
(D_\odot\rho^2_{\odot})^{-1} \int_{\ell =0}^\infty \rho^2[r(\ell,
\psi)] d\ell$; the coordinate $\ell$ runs along the line of sight,
which in turn makes an angle $\psi$ with respect to the direction of
the GC ( i.e., $r^2=\ell^2+D^2_\odot-2 \ell D_\odot \cos{\psi}$); the
subindex $f$ denotes the annihilation channels with one or more
photons in the final state and $dN_f/ dE_\gamma$ is the (normalized)
photon spectrum per annihilation; $\rho(\vec x)$, $\rho_\odot =
0.3~{\rm GeV}/{\rm cm}^3$, and $D_\odot \simeq 8.5~{\rm kpc}$
respectively denote the dark matter density at a generic location
$\vec x$ with respect to the GC, its value at the solar system
location, and the distance of the Sun from the GC.  In Fig.~\ref{fig3}
we show representative gamma ray spectra from dark matter
annihilations, assuming a dark matter distribution which follows the
Navarro-Frenk-White (NFW) halo profile~\cite{Navarro:1996gj}.  The
dotted line denotes the gamma ray spectrum from a 1~TeV neutralino
with a total annihilation rate $\sigma v|_{\rm tot} = 
10^{-9}~{\rm GeV}^{-2}$, but which annihilates to $\gamma \gamma$ or
$\gamma Z$ only 0.1\% of the time, which is typically for a TeV
neutralino in the MSSM. If the fraction of neutralino annihilations to
$\gamma \gamma$ were much larger, the prospects for detection would be
greatly improved. As previously noted, the stringy processes yield
much larger annihilation cross sections to this distinctive final
state. The solid line in Fig.~\ref{fig3}, corresponding to $\zeta =1$, shows the gamma ray spectrum
predicted for a neutralino which annihilates 10\% of the time to
$\gamma \gamma$. Unlike in the case of a typical MSSM neutralino, this
leads to a very bright and potentially observable gamma ray
feature. 

If an experiment were to detect a strong gamma ray line flux
without a corresponding continuum signal from the cascades of other
annihilation products, it could indicate the presence of a low string
scale.

\section{Direct detection experiments}

If the annihilation reaction is turned sideways, one can look at scattering for direct detection. In D-brane constructions the spin independent elastic scattering cross section is greatly supressed, because the $\chi$ is bouncing only on the gluon content of the nucleus,  
\begin{eqnarray}
\sigma_{\rm SI} & = & \frac{4}{\pi} m^2_p f^2_p \nonumber \\
       & = & \frac{4}{\pi} m^2_p \left(\frac{6}{9} \frac{1}{M_s^3} \pi  m_p N g_s^2 \wf f_{\rm TG}\right)^2 \nonumber \\
       & = & \frac{48}{27} \, \frac{1}{M_s^6} \, \pi \, m_p^4 \, N^2 g^4_s \, \wf \,  f_{\rm TG}^2  \,, 
\end{eqnarray}
where $f_p$ is the proton-WIMP coupling and $f_{\rm TG} \sim 0.83$ reflects the
gluon content of the nucleus~\cite{Jungman:1995df}. 
For our fiducial value $M_s \approx 2~{\rm TeV}$, we obtain $\sigma_{\rm SI} \approx 10^{-47}~{\rm cm}^2.$ This is well below the 90\%CL upper limits  $\sigma_{\rm SI} < 2 \times 10^{-43}~{\rm cm}^2$ and $\sigma_{\rm SI} < 3 \times 10^{-43}~{\rm cm}^2$  reported by the CDMS~\cite{Ahmed:2009zw} and XENON100~\cite{Aprile:2010um} collaborations, respectively.

\section{Conclusions}

Supersymmetric extensions of  D-brane
models can lead to an acceptable dark matter relic abundance of
bino-like neutralinos which annihilate a large fraction of the time
($\sim 10\%$) to $\gamma \gamma$, potentially producing a very bright
and distinctive gamma ray spectral line which could be observed by
current or next-generation gamma ray telescopes. Such a feature is
multiple orders of magnitude brighter than is typically predicted for
neutralino dark matter in the MSSM.

\bigskip 
\begin{acknowledgments}
I thank my collaborators Haim Goldberg, Dan Hooper, Danny Marfatia, and Tom Taylor for their many insights regarding the results summarized here. 
L.A.A.\ is supported by the U.S. National Science Foundation (NSF)
Grant No PHY-0757598, and the UWM Research Growth Initiative. 
\end{acknowledgments}

\bigskip 

\begin{thebibliography}{99} 




\bibitem{Bertone:2004pz}
  G.~Bertone, D.~Hooper and J.~Silk,
  Phys.\ Rept.\  {\bf 405}, 279 (2005)
  [arXiv:hep-ph/0404175];
  J.~L.~Feng,
  Annals Phys.\  {\bf 315}, 2 (2005)
 [arXiv:hep-ph/0405215]. 


\bibitem{Jungman:1995df}
  G.~Jungman, M.~Kamionkowski and K.~Griest,
  Phys.\ Rept.\  {\bf 267}, 195 (1996)
  [arXiv:hep-ph/9506380].

\bibitem{Komatsu:2008hk}
  E.~Komatsu {\it et al.}  [WMAP Collaboration],
  Astrophys.\ J.\ Suppl.\  {\bf 180}, 330 (2009)
  [arXiv:0803.0547 [astro-ph]].



\bibitem{Navarro:1996gj}
  J.~F.~Navarro, C.~S.~Frenk and S.~D.~M.~White,
  Astrophys.\ J.\  {\bf 490}, 493 (1997)
  [arXiv:astro-ph/9611107].

\bibitem{Aharonian:2004wa}
  F.~Aharonian {\it et al.}  [H.E.S.S. Collaboration],
  Astron.\ Astrophys.\  {\bf 425}, L13 (2004)
  [arXiv:astro-ph/0408145];
  F.~Aharonian {\it et al.}  [H.E.S.S. Collaboration],
  Phys.\ Rev.\ Lett.\  {\bf 97}, 221102 (2006)
  [Erratum-ibid.\  {\bf 97}, 249901 (2006)]
  [arXiv:astro-ph/0610509].






\bibitem{Goldberg:1983nd}
  H.~Goldberg,
  Phys.\ Rev.\ Lett.\  {\bf 50}, 1419 (1983);
  J.~R.~Ellis, J.~S.~Hagelin, D.~V.~Nanopoulos, K.~A.~Olive and M.~Srednicki,
  Nucl.\ Phys.\  B {\bf 238}, 453 (1984).

\bibitem{Wess-Bagger} J. Wess and J. Bagger, {\it Supersymmetry and Supergravity}, (Princeton University Press, 1992) p.33.


\bibitem{Bergstrom:1997fh}
  L.~Bergstrom and P.~Ullio,
  Nucl.\ Phys.\  B {\bf 504}, 27 (1997)
  [arXiv:hep-ph/9706232];
  Z.~Bern, P.~Gondolo and M.~Perelstein,
  Phys.\ Lett.\  B {\bf 411}, 86 (1997)
  [arXiv:hep-ph/9706538];
  P.~Ullio and L.~Bergstrom,
  Phys.\ Rev.\  D {\bf 57}, 1962 (1998)
  [arXiv:hep-ph/9707333].




\bibitem{Blumenhagen:2006ci} 
  R.~Blumenhagen, B.~K\"ors, D.~L\"ust and S.~Stieberger,
  Phys.\ Rept.\  {\bf 445}, 1 (2007)
  [arXiv:hep-th/0610327].



\bibitem{Antoniadis:2000ena}
  I.~Antoniadis, E.~Kiritsis and T.~N.~Tomaras,
  Phys.\ Lett.\  B {\bf 486}, 186 (2000)
  [arXiv:hep-ph/0004214].


  \bibitem{Antoniadis:2004qn}
  I.~Antoniadis and T.~R.~Taylor,
  Nucl.\ Phys.\  B {\bf 695}, 103 (2004)
  [arXiv:hep-th/0403293];
  I.~Antoniadis and T.~R.~Taylor,
  Nucl.\ Phys.\  B {\bf 731}, 164 (2005)
  [arXiv:hep-th/0509048].

\bibitem{Antoniadis:2005sd}
  I.~Antoniadis, K.~S.~Narain and T.~R.~Taylor,
  Nucl.\ Phys.\  B {\bf 729}, 235 (2005)
  [arXiv:hep-th/0507244].



\bibitem{Anchordoqui:2009bn}
  L.~A.~Anchordoqui, H.~Goldberg, D.~Hooper, D.~Marfatia and T.~R.~Taylor,
  Phys.\ Lett.\  B {\bf 683}, 321 (2010)
  [arXiv:0912.0517 [hep-ph]].


\bibitem{Anchordoqui:2007da}
  L.~A.~Anchordoqui, H.~Goldberg, S.~Nawata and T.~R.~Taylor,
  Phys.\ Rev.\ Lett.\  {\bf 100}, 171603 (2008)
  [arXiv:0712.0386 [hep-ph]];
  L.~A.~Anchordoqui, H.~Goldberg, D.~Lust, S.~Nawata, S.~Stieberger and T.~R.~Taylor,
  Phys.\ Rev.\ Lett.\  {\bf 101}, 241803 (2008)
  [arXiv:0808.0497 [hep-ph]].

\bibitem{Bergstrom:1997fj}
  L.~Bergstrom, P.~Ullio and J.~H.~Buckley,
  Astropart.\ Phys.\  {\bf 9}, 137 (1998)
  [arXiv:astro-ph/9712318].

\bibitem{Aharonian:2004jr}
  F.~Aharonian and A.~Neronov,
  Astrophys.\ J.\  {\bf 619}, 306 (2005)
  [arXiv:astro-ph/0408303];
  A.~Atoyan and C.~D.~Dermer,
  Astrophys.\ J.\  {\bf 617}, L123 (2004)
  [arXiv:astro-ph/0410243].



\bibitem{Ahmed:2009zw}
  Z.~Ahmed {\it et al.}  [The CDMS-II Collaboration],
Science {\bf 327}, 1620 (2010)   
[arXiv:0912.3592 [astro-ph.CO]].


\bibitem{Aprile:2010um}
  E.~Aprile {\it et al.}  [XENON100 Collaboration],
  arXiv:1005.0380 [astro-ph.CO].

\end{thebibliography}

\end{document}